\documentclass[final,5p,times,twocolumn]{elsarticle}
\usepackage{graphicx}

\begin{document}

	\begin{frontmatter}
\title{Hubbard bands and exotic states in doped and undoped Mott systems: the Kotliar-Ruckenstein representation
}



\author{V. Yu. Irkhin
\/\thanks{e-mail: valentin.irkhin@imp.uran.ru}
}



\address{M. N. Mikheev Institute of Metal Physics, 620108 Ekaterinburg, Russia}

\begin{abstract}
{
The slave-particle representation  is promising	
method to treat the properties of exotic strongly correlated systems. We develop a unified approach to describe both paramagnetic state with possible spin-liquid features and states with strong long-range or short-range magnetic order.
 Combining the Kotliar-Ruckenstein representation and fractionalized spin-liquid deconfinement picture,  the Mott transition and  Hubbard subbands are considered.   The spectrum in the insulating state is significantly affected by the presence of the spinon spin-liquid spectrum and a hidden Fermi surface. Presenting a modification of the Kotliar-Ruckenstein representation in the spin-wave region, we  treat the case of magnetic order, especial attention being paid to half-metallic ferromagnetic state. The formation of  small and large Fermi surfaces for doped current carriers in the antiferromagnetic state is also discussed.
}
\end{abstract}
\end{frontmatter}



\section{Introduction}

The problem of describing strongly correlated states has been a topic of interest and significance for a long time. In particular, here belong the aspects of the Mott transition, which refers to the correlation-driven transition from a metallic state to an insulating state \cite{Mott}. The related physical phenomena occur in a number of doped and undoped Mott systems, including  insulators and metals with exotic properties \cite{Imada}.

The physics of the Mott systems   originates from competition of magnetism, Coulomb correlations, frustration and topology.
Typically (in most d-metal compounds), the Mott transition occurs  according to the Slater mechanism, i.e., involves the  insulating phase  with antiferromagnetic band splitting (see, e.g., Ref. \cite{Igoshev}). However, the situation changes when dealing with  frustrated systems which do not demonstrate antiferromagnetic ordering, so that only the paramagnetic metallic and insulator states (possibly, with unusual characteristics) are present, leading to the formation of a spin-liquid-type state \cite{Vojta,Senthil2}.

Such a  transition into the insulator state, known as the Mott scenario, is associated with the correlation Hubbard splitting. In the Mott state, the spectrum exhibits a significant charge gap that is determined by bosonic excitation branches.  Consequently, the electrons become composite particles and undergo fractionalization, where the spin characteristics are controlled by neutral fermions called spinons, and the charge ones  by bosons \cite{Anderson,Wen}. This concept can be formalized by using the slave-boson representations
\cite{Kotliar86,Anderson,Wen}.

The interaction between bosons and fermions mediated by a gauge field plays a significant role as it gives rise to confinement \cite{Wen}. This leads to a transition towards a confinement metallic state, which is marked by the occurrence of Bose condensation and a non-zero residue in the electron Green's function. Conversely, in the insulator state, the bosons have a gap in their energy spectrum, leading to an incoherent overall spectrum that encompasses Hubbard's bands. In this case, the electron Green's function is a combination of the boson and fermion Green's functions through convolution.


Recent theoretical advancements have offered a fresh perspective on the Mott transition by introducing a topological framework. This is particularly relevant because spin liquids, known for their topological order, are involved in this transition. In the study of phase transitions in magnetically frustrated systems,  the consideration of topological excitations  becomes essential as they play a significant role in confinement. These ideas have been extensively reviewed, e.g., in Refs. \cite{Wen22,Scr2}.



As for doped Mott systems, copper-oxide  materials which are basic for high-$T_c$
 superconductors should be mentioned to the first place. In the overdoped case, the 
normal (non-superconducting)  ground state 
 is characterized as a Fermi liquid
with a ``large'' Fermi surface (including both localized and itinerant states), where Luttinger's theorem holds. At the same time, in the underdoped case the ground state is more complicated and may possess small hole pockets of the Fermi surface \cite{Wen,Senthil1}. The description of this state depends again on the presence of absence of antiferromagnetic ordering.
The small Fermi surface can occur not only in the case of long-range order,  but also in the situation of  strong short-range order \cite{Punk1,Punk,Sach}.

In this paper, we examine the metal-insulator transition through a topological perspective, specifically focusing on spin-charge separation within the framework of the Kotliar-Ruckenstein slave-boson representation. We employ the deconfinement concept  to investigate the Hubbard subbands' spectrum. Our treatment aims to understand the Mott transition leading to a spin-liquid state, while also establishing the connection between the charge gap in the boson spectrum and the Hubbard splitting.

The idea of preserving the Fermi surface during a quantum phase transition is supported by the presence of a spinon Fermi surface in the paramagnetic phase of a Mott insulator \cite{Senthil2}. In a gapped phase like the Mott state, the traditional Fermi surface does not exist and instead transforms into a hidden or ghost Fermi surface. However, the volume enclosed by the Fermi surface, as described by the Luttinger theorem, remains conserved \cite{Volovik}. This concept has also been applied to half-metallic ferromagnets \cite{Irk, Scr3}. In this study, we expand upon this approach and demonstrate how to combine the concept of composite particles with spin-liquid states and magnetic ordering in various cases.

In Sec. 2 we review various versions of the slave-boson representations. In Sec. 3 we treat the problem of metal-insulator  in the paramagnetic case. Although we apply the standard Kotliar-Ruckenstein representation used in previous works \cite{lavagna,raimondi}, we provide a new interpretation   which takes into account  spin-charge separation in terms of exotic quasiparticles -- spinons and holons. In Sec. 4 we derive a new form of the Kotliar-Ruckenstein representation, which is compatible with the approach of many-electron Hubbard's operators \cite{II} and is convenient in the magnetic state.  We  apply this form to treat conducting ferromagnets and antiferromagnets.  In Sec. 5, a discussion is presented.


\section{Slave-particle representations of the Hubbard model}

The Hamiltonian of the Hubbard model reads
\begin{equation}  \label{eq:original_H}
	\mathcal{H} = -\sum_{ij\sigma}
	t_{ij}c^\dag_{i\sigma}c^{}_{j\sigma}+U\sum_i n_{i\uparrow}n_{i\downarrow}+\mathcal{H}_d,
\end{equation}
where $c^{\dagger}_{i\sigma}$ are electron creation operators.
The Heisenberg interaction
\begin{equation} 
{H}_d=\sum_{ij }J_{ij}{\bf S}_i{\bf S}_j, 
\end{equation}
which can arise as an effective superexchange interaction in the second order of perturbation theory in the Hubbard model, is explicitly incorporated 
for further ease of representation. 
 Such a mixed representation is known as $t-J-U$ model which reduces in the  large-$U$ limit to the well-known $t-J$ model (see, e.g., the review \cite{Spa}).
The Hamiltonian of the latter model for the hole doping can be represented in the form
\begin{equation}  
	\mathcal{H} = \sum_{ij\sigma}	t_{ij}\tilde{c}^\dag_{i\sigma}\tilde{c}^{}_{j\sigma}+\mathcal{H}_d \label{H2}
\end{equation}
where  $\tilde{c}^\dag_{i\sigma}=X_i(0\sigma)=|i0\rangle\langle i\sigma|={c}_{i \sigma}(1-n_{i -\sigma })$ are the Hubbard projection operators creating empty on-site states.

In situations where strong correlation effects are dominant, it is often useful to employ auxiliary or ``slave'' boson and fermion representations. 
The slave-boson representation was proposed in the pioneering works by Barnes \cite{Barnes} and Coleman \cite{Coleman} for the Anderson models and developed by many authors.

Anderson \cite{Anderson} proposed a physical interpretation of slave-boson representation for the Hubbard model based on the concept of separating the spin and charge  degrees of freedom  of an electron,
\begin{equation}
	c_{i\sigma }=X_i(0,\sigma )+\sigma X_i(-\sigma,2)=e_i^{\dagger }f_{i\sigma
	}+\sigma d_i f^{\dagger }_{i-\sigma }.  \label{eq:6.131}
\end{equation}
where $\sigma = \pm 1$, $f_{i\sigma }$ are the annihilation
operators for neutral fermions (spinons), and $e_i$, $d_i$ for charged spinless bosons (holons and doublons).
In the large-$U$ limit we have to retain in (\ref{eq:6.131}) only the first (second) term for the hole (electron) doping.

Alternatively,  the slave-fermion representation which uses the Schwinger boson operators $b_{i\sigma }$ can be used (see, e.g., Ref. \cite{Kane}),
\begin{equation}
X_{i}(0,\sigma )=f_{i}^{\dagger}b_{i\sigma },\, X_{i}(+,- )= b^\dag_{i\uparrow }b_{i\downarrow },  \label{eq:6.31}
\end{equation}%
 so that
\begin{equation}
 \sum_{\sigma}b^\dag_{i\sigma }b_{i\sigma }+f_{i}^{\dagger}f_{i}=1.  \label{eq:6.122}
\end{equation}%
This representation is more suitable in the case of magnetic ordering, Such uncertainty in statistics of excitations leads to difficulties in constructing a unified picture and requires more advanced approaches.


A more complicated representation was proposed by Kotliar and Ruckenstein~\cite{Kotliar86}.
This  uses the Bose operators $
e_{i},\,p_{i\sigma },\,d_{i}$ and Fermi  operators $f_{i\sigma }$: 
\begin{equation}
	c_{i\sigma }^{\dag }\rightarrow f_{i\sigma }^{\dag }z_{i\sigma }^{\dag
	},~z_{i\sigma }^{\dag }=g_{2i\sigma }(p_{i\sigma }^{\dag
	}e_{i}^{{}}+d_{i}^{\dag }p_{i-\sigma }^{{}})g_{1i\sigma }, \label{KR}
\end{equation}%
with the constraints
\begin{equation}
	\sum_{\sigma }p_{i\sigma }^{\dagger }p_{i\sigma }+e_{i}^{\dagger
	}e_{i}+d_{i}^{\dagger }d_{i}=1,~\,f_{i\sigma }^{\dagger }f_{i\sigma
	}=p_{i\sigma }^{\dagger }p_{i\sigma }+d_{i}^{\dagger }d_{i}  \label{con},
\end{equation}%
which can be used to introduce gauge fields \cite{Wen}.

According to Kotliar and Ruckenstein, the representation of many-electron operators is not fixed and can include additional operator factors as long as they have eigenvalues of 1 in the physical subspace. While all these forms yield accurate results in exact treatments, they may differ in approximate calculations. This is particularly significant when constructing mean-field approximations as it allows for agreement with limiting cases.
Thus  the factors  $g_{1,2i\sigma}$ are somewhat arbitrary, but to obtain an agreement with the Hartree-Fock limit one uses the values

\begin{eqnarray}
g_{1i\sigma } =(1-d_{i}^{\dag }d_{i}-p_{i\sigma }^{\dag }p_{i\sigma })^{-1/2},   \nonumber \\
g_{2i\sigma }=(1-e_{i}^{\dag }e_{i}-p_{i-\sigma}^{\dag }p_{i-\sigma})^{-1/2} .
\label{zzz}
\end{eqnarray}

In the mean-field approximation for a non-doped case and a non-magnetic state we can put $g_{1,2\sigma}^{2}=2$. 
It should be noted that such a choice results in some difficulties, in particular leads to inconsistency in the atomic limit \cite{raimondi}.
Also, we will see below that this choice is inadequate
in a magnetic state.

In the framework of various slave-boson approaches, a number of mean-field theories were developed \cite{Wen}. In particular, treatments within the Kotliar and Ruckenstein  representations on saddle-point level became popular because of their good agreement with numerical simulations.  However, such treatments are not free of difficulties \cite{Jolicoeur,Arrigoni}. Generally speaking, they suffer from drawbacks connected with  spurious Bose condensation. To overcome this difficulty and develop more advanced theories, one can use the $1/N$-expansion \cite{Coleman} or gauge-field theories which are extensively discussed in the review \cite{Wen}.
In this connection,  treatments of the limiting cases, where the slave-boson approach is exact or  controlled \cite{Kopp1,Kopp2}, can be useful.

To take into account properly spin-flip processes, it is suitable to use the rotationally invariant version \cite{Li,Fresard:1992a}.
Here projected electron is represented as a composite of Fermi spinon with
scalar and vector bosons $p_{i0}$ and $\mathbf{p}_{i}$. Using the coupling
rule of momenta 1 and 1/2 one obtains
\begin{equation}
c_{i\sigma }=\sum_{\sigma ^{\prime }}(e_{i}^{\dag }p_{i\sigma ^{\prime
}\sigma }+\sigma p_{i-\sigma -\sigma ^{\prime }}^{\dag }d_{i})f_{i\sigma
^{\prime }} \label{KR1}
\end{equation}%
with
\begin{equation}
\hat{p}_{i}=\frac{1}{2}(p_{i0}\sigma _{0}+\mathbf{p}_{i}\mbox{\boldmath$%
\sigma $})
\end{equation}%
%
and the constraints
\begin{equation}
e_{i}^{\dag }e_{i}^{{}}+\sum_{\mu =0}^{3}p_{i\mu }^{\dag }p_{i\mu
}^{{}}+d_{i}^{\dag }d_{i}^{{}}=1,  \label{sum}
\end{equation}%

\begin{equation}
\sum_{\sigma}f^{\dag}_{i\sigma}f_{i\sigma}=\sum_{\mu =0}^{3}p_{i\mu }^{\dag }p_{i\mu
}^{{}}+2d_{i}^{\dag }d_{i}^{{}}.  \label{sum2}
\end{equation}%
Introducing proper factors one has \cite{Fresard:1992a}
\begin{equation}
	c_{i\sigma }=\sum_{\sigma ^{\prime }}f_{i\sigma ^{\prime }}z_{i\sigma
		^{\prime }\sigma },~\hat{z}_{i}=e_{i}^{\dag }\hat{L}_{i}M_{i}\hat{R}_{i}\hat{%
		p}_{i}+\widehat{\tilde{p}}_{i}^{\dag }\hat{L}_{i}M_{i}\hat{R}_{\hat{\imath}%
	}d_{i}  \label{zz}
\end{equation}%
where
\begin{eqnarray}
	\hat{L}_{i} &=&[(1-d_{i}^{\dag }d_{i})\sigma _{0}-2\widehat{p}_{i}^{\dag }%
	\widehat{p}_{i}]^{-1/2}  \label{1z1} \\
	\hat{R}_{i} &=&[(1-e_{i}^{\dag }e_{i})\sigma _{0}-2\widehat{\tilde{p}}%
	_{i}^{\dag }\widehat{\tilde{p}}_{i}]^{-1/2}  \label{1z2} \\
	M_{i} &=&(1+e_{i}^{\dag }e_{i}^{{}}+\sum_{\mu =0}^{3}p_{i\mu }^{\dag
	}p_{i\mu }^{{}}+d_{i}^{\dag }d_{i}^{{}})^{1/2}.  \label{1z}
\end{eqnarray}%
The additional square-root factors in (\ref{1z1})-(\ref{1z}) can be treated
in spirit of mean-field approximation. In particular, the factor $M$ 
is equal to $\sqrt{2}$ due to sum rule (\ref{sum}) and enables one to obtain an agreement with the small-$U$ limit and with
the saturated ferromagnetic case.
The scalar and vector bosons $p_{i0}$ and $\mathbf{p}_{i}$ are introduced as
\begin{equation}
	\hat{p}_{i}=\frac{1}{2}(p_{i0}\sigma _{0}+\mathbf{p}_{i}\mbox{\boldmath$%
		\sigma $})
\end{equation}%
with $\mbox{\boldmath$\sigma $}$ being Pauli matrices and $\hat{\tilde{p}}%
_{i}=(1/2)(p_{i0}\sigma _{0}-\mathbf{p}_{i}\mbox{\boldmath$\sigma $})$ the
time reverse of operator $\hat{p}_{i}$. 

 In Sec. 4,  we will extensively employ the rotationally invariant representation to treat in detail the magnetically ordered case. We will perform the corresponding analytical transformations and demonstrate that the full form of the radicals  plays an important role. In particular,  this is crucial to describe incoherent states in a ferromagnet.

\section{Mott transition and Hubbard bands in the paramagnetic and spin-liquid state}

In order to treat the Mott transition in frustrated systems within the paramagnetic phase, several studies \cite{Senthil2,Lee,Hermele1} utilized the rotor representation. While this representation is straightforward, it is not ideal as it does not explicitly incorporate the spectrum of both Hubbard bands.
An alternative  description of the   Mott transition and Hubbard bands  can be obtained within the Kotliar-Ruckenstein representation  \cite{lavagna,raimondi}
These works use a Gutzwiller-type approach for a structureless  paramagnetic state. Here we perform a more advanced treatment with account of possible spin-liquid picture.
To take into account spin frustrations, we include explicitly into the model the Heisenberg interaction.
Then the  Lagrangian of the Hubbard-Heisenberg model has the form
\begin{eqnarray}
	\mathcal{L} &=&-\sum_{ij\sigma }t_{ij}z_{i\sigma }^{\dag }z_{j\sigma
	}f_{i\sigma }^{\dag }f_{j\sigma }+\sum_{i\sigma }f_{i\sigma }^{\dag
	}(\partial _{\tau }-\mu +\lambda _{2\sigma })f_{i\sigma }  \nonumber \\
	&+&\sum_{i\sigma }p_{i\sigma }^{\dag }(\partial _{\tau }+\lambda
	_{1}-\lambda _{2\sigma })p_{j\sigma }+\sum_{i}e_{i}^{\dag }(\partial _{\tau
	}+\lambda _{1})e_{i}  \nonumber \\
	&+&\sum_{i}d_{i}^{\dag }(\partial _{\tau }+\lambda _{1}-\sum_{\sigma
	}\lambda _{2\sigma }+U)d_{i}+\mathcal{H}_d.  \label{lag}
\end{eqnarray}%
By employing the Heisenberg Hamiltonian in the $f$-pseudofermion representation, it is possible to analyze spin degrees of freedom independently. In certain circumstances, it is anticipated that a spin-liquid state may emerge, characterized by excitations primarily consisting of spinons, which are neutral fermions.

In the mean-field approximation, the Lagrange factors $\lambda_{1,2}$ associated with (\ref{con}) are not dependent on the specific sites. When in the insulator phase, it has been established by Lavagna  \cite{lavagna} that $\lambda _{1 }=\lambda _{2 \sigma} = U(1 \pm\zeta)/2$ equals to the chemical potential for a infinitesimally small electron or hole doping (the addition or removal of an electron),
  $\zeta =(1-1/u)^{1/2}$, $u=U/U_c$. Here 
$$U_{c}=4p^2 g_{1}^{2}g_{2}^{2}\varepsilon=8 \varepsilon $$ 
is the critical value for the Mott transition
in the Brinkman-Rice approximation (see Ref. \cite{Vollhardt}),  $\varepsilon =2\left\vert \int_{-\infty }^{\mu }d\omega \omega \rho
(\omega )\right\vert $
the average energy of non-interacting electron system, $\rho (\omega )$ the bare density of
 states. 

Following to Refs.\cite{Vollhardt,lavagna} we can introduce the variable $x=e+d.$ Then we obtain for $y=1/x^{2}$ the cubic equation 
\begin{equation}
	y^{3}-(u-1)y/u\delta ^{2}=1/u\delta ^{2}.
\end{equation}%


Earlier the solution of this  equation was discussed in Refs. \cite{lavagna,Fresard:1992a}. Here we present the solution in a more convenient form.
Passing to the variable $1/y$ and using trigonometric solution of the cubic equation we derive for $u<1$ (correlated metal phase) 
\begin{equation}
	y=2\left( \frac{1-u}{3u\delta ^{2}}\right) ^{1/2}\sinh \left( \frac{1}{3}{\rm arcsinh}%
	\frac{\delta }{\delta _{0}}\right) 
\end{equation}%
For $u>1$ one has
\begin{equation}
	y=2\left( \frac{u-1}{3u\delta ^{2}}\right) ^{1/2}\times \left\{ 
	\begin{array}{cc}
		\cos \left( \frac{1}{3}{\rm arccos}(\delta /\delta _{0}) \right),  & \delta
		<\delta _{0} \\ 
		\cosh \left( \frac{1}{3}{\rm arccosh}(\delta /\delta _{0})\right) , & \delta
		>\delta _{0}%
	\end{array}%
	\right. 
\end{equation}%
where 
\begin{equation}
	\delta _{0}=2|u-1|^{3/2}/(27u)^{1/2}  \label{d2}
\end{equation}%
This solution is a smooth and analytic function of doping $\delta $ in the whole
region $\delta <1.$ For small $\delta \ll \delta _{0}$ we have 
\begin{equation}
	x^{2}=1/y=\left\{ 
	\begin{array}{cc}
		1-u +O(\delta ^{2}), & ~u<1 \\ 
		\delta /\sqrt{1-1/u}, & u>1%
	\end{array}%
	\right.   \label{n1}
\end{equation}
Generally, a considerable  $U$-dependence takes place at any $\delta .$ For 
$\delta \gg \delta _{0}$ (close to the Mott transition) we have $x^{2}\simeq \delta^{2/3}.$


The behavior (\ref{n1}) can be considerably changed when taking into account gauge fluctuations \cite{Senthil2,Senthil3}, especially in the two-dimensional case where intermediate energy and  temperature  scales can occur beyond mean-field picture.

It is convenient to introduce the boson combination  $b_{i}^{\dag }=e_{i}^{\dag }+d_{i}$ yields (cf. Ref.\cite{raimondi}). The expression of the corresponding Green's function takes the form
\begin{eqnarray}
	D(\mathbf{q},\omega ) &=&\langle \langle b_{\mathbf{q}}|b_{\mathbf{q}}^{\dag
	}\rangle \rangle _{\omega }=\sum_{a =1,2}\frac{Z_{\alpha \mathbf{q}}}{		\omega -\omega _{\alpha \mathbf{q}}},~ \\
	Z_{a \mathbf{q}} &=&(-1)^{a}U/\sqrt{U^{2}\zeta ^{2}+U (U_{c}- 4\Sigma ({\mathbf{q}}))}
\end{eqnarray}
with the spectrum of boson subsystem 
\begin{eqnarray}
	\omega _{a \mathbf{q}} &=&\frac{1}{2}[\pm U\zeta - (-1)^a \sqrt{U^{2}\zeta		^{2}+U (U_{c}- 4 \Sigma ({\mathbf{q}}))}] \label{la}
\end{eqnarray}%
One of two boson branches becomes gapless and provides formation of the boson condensate at the Mott transition.

To obtain the boson self-energy  we perform a decoupling of the first term in (\ref{lag}), which yields essentially the correlation correction first introduced in Ref.\cite{Harris}. 
The result reads
\begin{equation}
	\Sigma (\mathbf{q})=-p^2 g_{1}^{2}g_{2}^{2}\sum_{\mathbf{k}\sigma }t_{\mathbf{k-q}}n_{\mathbf{k}\sigma },\, n_{\mathbf{k}\sigma}=\langle f_{\mathbf{k}\sigma}^{\dag }f_{\mathbf{k}\sigma}^{ }\rangle . \label{SF}
\end{equation}%
In Ref. \cite{raimondi},  the limit of vanishing renormalized electron bandwidth (i.e., bearing in mind the Mott phase where the averages $e,\,d\rightarrow 0$) was treated in a Gutzwiller-type approach. Here we use a more straightforward approach: a finite bandwidth of holons occurs in a natural way by taking into account the spinon dispersion. Note that earlier a similar consideration was performed for the $t-J$ model \cite{Wen}.

The presence of a small (as compared to electron energies) characteristic scale of spinon energies  is crucial. As a result, the temperature dependence of the spinon Fermi surface becomes significant. This scenario shares similarities with the situation observed in magnetic order (e.g., band spliting owing to long- or short-range antiferromagnetic ordering).
The dispersion of bosons is affected by the specific characteristics of the fermion spectrum, which are determined by the state of the $f$-system.


The spinon spectrum $E_{\bf k}$ can be stabilized in the mean-field scenario through either a non-compact gauge field or by having gapless Fermi excitations \cite{Nagaosa1,Senthil2,I23}. In the insulator state, this spectrum remains unaffected by bosons, leading to the emergence of various spin-liquid phases \cite{Wen}.


When there is minimal dependence on $\mathbf{k}$ of $n_{\mathbf{k}\sigma}$ (indicating a localized spin phase without fermion hopping), the value of $\Sigma$ approaches zero. However, in the case of a spin liquid, a distinct Fermi surface is present. Despite the spectrum of spinons can differ from that of bare electrons, putting $q=0$ we still obtain $\Sigma(0)=U_c/4$, since the spinon band is half-filled and the position of the Fermi energy (the chemical potential) remains fixed.

In the nearest-neighbor approximation, when converting equation (\ref{SF}) into real-space representation, it becomes evident that the spinon spectrum and the correction to the holon spectrum vary only in terms of replacing the parameter $J$ with $t$ ($\Sigma({\bf q}) \propto E(\bf{q})$, as described in Ref. \cite{Wen} for the $t-J$ model). Specifically, we can observe that
\begin{eqnarray}
	\Sigma (\mathbf{q})=U_c(\cos q_x+\cos q_y)/8, \nonumber \\
	\Sigma (\mathbf{q})=\pm U_c\sqrt{\cos^2 q_x + \cos^2 q_y}/(4\sqrt{2})
\end{eqnarray}%
for  Anderson's uniform resonating valence bond  (uRVB) and $\pi$-flux ($\pi$Fl)  phases, respectively.
Thus in the case of uRVB state the quasimomentum dependences of electron and spinon spectrum coincide: 
$E_{\bf k} \sim J (\cos k_x+\cos k_y)$. 
At the same time,  our method enables one to treat a more general situation.
So, in the $\pi$Fl phase (which includes Dirac points) $E_{\bf k} \sim \pm J \sqrt{\cos^2 k_x + \cos^2 k_y}$. 
For the gapped Z$_2$ phase, which can occur in the presence of next-nearest-neighbor interactions, the mapping of the spectra is violated and the consideration is more difficult.


In the case of large $U$  we have two well-separated bands $$\omega_{a\mathbf{q}}= {\rm const} - (-1)^a \Sigma({\bf q})/\zeta.$$

The observable electron Green's function is obtained as a convolution of the boson and spinon Green's functions \cite{raimondi,Wen,I23}.
For $J \ll  |t|$, this spinon smearing 
does not strongly influence the shape of density of state. Then we can put ${\rm Im} \langle \langle f_{\mathbf{k\sigma }}^{ }|f_{\mathbf{k\sigma }}^{\dag }\rangle \rangle _{E} \sim \delta(E-\lambda_2)$ to obtain the Hubbard bands with the energies $\lambda_2 - \omega_{1,2 \mathbf{q}}$ for vanishing electron (hole) doping with energies near 0 and $U$, respectively, $\lambda_2 $ being the corresponding chemical potential \cite{raimondi}.
This energy spectrum consists of upper and lower Hubbard subbands, each with a width of order of the bare bandwidth. At the transition point where the interaction strength  approaches the critical value $U_c$, the energy gap between these subbands diminishes and eventually closes. A further analysis of collective modes arising from the Hubbard bands with account of doping was performed in Ref. \cite{Dao}.

\section{Magnetic states of the doped Mott insulator}

\subsection{Derivation of the Hamiltonian in the spin-wave region}

For magnetically ordered phase with strong long-range or short-range order the  approximations of previous section  do not work since the above approximation for factors $g$  is not valid \cite{Irk}.
Most simple is the case of ferromagnetic ordering which was investigated earlier in terms of Hubbard's operators \cite{FTT,FTT1}.
According to Nagaoka \cite{Nagaoka}, the ground state in the large-$U$ limit is a saturated ferromagnet for one excess hole (or double); this conclusion can be extended on the case of finite doping, as demonstrated from analysis of instabilities of this state which can be characterized as a half-metallic ferromagnet with an energy gap for one of spin projections \cite{FTT,FTT1}. 

The original version of the Kotliar-Ruckenstein	 representation (\ref{KR}) provides a mean-field description, but turns out to be insufficient, since it does not describe spin-flip processes which are crucial to describe incoherent states. Therefore we use the rotationally invariant representation (\ref{KR1}) and carry out its further transformations.

The square-root factors in (\ref{1z}) can be treated in spirit of mean-field approximation. Correspondingly,  the factor $M$
is put $\sqrt{2}$ due to sum rule (\ref{sum}); this permits to obtain an agreement with the free-electron limit and with the ferromagnetic case.




According to Ref. \cite{Fresard:1992a}, we have
\begin{equation}
\mathbf{S}=\frac{1}{2}\sum_{{}}\mbox{\boldmath$\sigma $}_{\sigma \sigma
^{\prime }}p_{\sigma \sigma _{1}}^{\dag }p_{\sigma _{2}\sigma ^{\prime }}=%
\frac{1}{2}(p_{0}^{\dag }\overline{\mathbf{p}}+\overline{\mathbf{p}}^{\dag
}p_{0}-i[\overline{\mathbf{p}}^{\dag }\times \overline{\mathbf{p}}])
\label{SS}
\end{equation}%
with $\overline{\mathbf{p}}=(p^{x},-p^{y},p^{z})$.
Then we derive
\begin{eqnarray}
S^{z} &=&\frac{1}{2}(p_{0}^{\dag }p_{z}+p_{z}^{\dag }p_{0}+i(p_{x}^{\dag 
}p_{y}-p_{y}^{\dag }p_{x}))  \nonumber\\
&=&\frac{1}{2}(p_{0}^{\dag }p_{z}+p_{z}^{\dag }p_{0}+p_{{}}^{+\dag 
}p_{{}}^{+}-p_{{}}^{-\dag }p_{{}}^{-}) \nonumber \\
&=&\frac{1}{2}(1-(p_{0}^{\dag }-p_{z}^{\dag })(p_{0}-p_{z}))-p_{{}}^{-\dag 
}p_{{}}^{-} \label{s1} \\
S^{+} &=&\frac{1}{\sqrt{2}}((p_{0}^{\dag }+p_{z}^{\dag })p^{-}+p^{+\dag
}(p_{0}^{{}}-p_{z})) \label{s2}
\end{eqnarray}%
where $p^{\pm }=(p_{x}^{{}}\pm ip_{y})/\sqrt{2}$ and we have taken into
account (\ref{sum}). One can see that commutation relations for spin
operators are exactly satisfied, unlike the linearized Holstein-Primakoff
representation.

For a Heisenberg ferromagnet  ($p_0\simeq p^z \simeq 2^{-1/2}$)
we obtain $S_{i}^{+}\simeq p_{i}^{-}$ to lowest-order approximation, 
so that the Heisenberg Hamiltonian takes the usual spin-wave form
\begin{equation}
\mathcal{H}_{d}=\sum_{\mathbf{q}}\omega _{\mathbf{q}}p_{\mathbf{q}%
}^{-}{}^{\dag }p_{\mathbf{q}}^{-}+\mathrm{const},\,\omega _{\mathbf{q}}=J_{%
\mathbf{q}}-J_{0}
\end{equation}


It is crucial to highlight that in order to achieve this outcome, it is essential to retain the vector product in equation (\ref{SS}) to prevent mixing of the bosons $p$ and $p^{\dag}$. This retainment differs from the approach employed in Ref. \cite{Fresard:1992a} for the paramagnetic phase. Note that in the magnetic ordering case $p_{i}^{+}$ is not related to spin operators, see (\ref{s1}), (\ref{s2}).

Eq.(\ref{zz}) can be simplified in the case of half-metallic ferromagnetism near half-filling (small doping, band filling $n\lesssim 1$) where, in the mean-field approach, $%
p_{0}=p_{z}=p\simeq 1/\sqrt{2}$, $e\simeq \langle e\rangle =(1-n)^{1/2}$.
Taking into account the relation
\begin{equation}
2\widehat{p}_{i}^{\dag }\widehat{p}_{i}=\frac{1}{2}(p_{i0}^{2}\sigma _{0}+(%
\mathbf{S}_{i}\mbox{\boldmath$\sigma $}))
\end{equation}%
we obtain 
\begin{eqnarray}
~L &=&\left(
\begin{array}{cc}
1-p_{0}^{2}-S^{z} & -S^{+} \\
-S^{-} & 1-p_{0}^{2}+S^{z}%
\end{array}%
\right)^{-1/2} , \\
R &=&\left(
\begin{array}{cc}
1-p_{0}^{2}+S^{z}-e_{i}^{\dag }e_{i} & -S^{-} \\
-S^{+} & 1-p_{0}^{2}-S^{z}-e_{i}^{\dag }e_{i}%
\end{array}%
\right) ^{-1/2}, \\
p &=&\left(
\begin{array}{cc}
p_{0}+p_{z} & p^{-} \\
p^{+} & p_{0}-p_{z}%
\end{array}%
\right)
\end{eqnarray}%
Using the sum rule (\ref{sum}) and retaining only diagonal terms we obtain $%
L_{++}\sim 1/|e|,$ $R_{--}\sim 1/|p^{\pm }|.$
Neglecting the terms proportional to holon operators we derive in the large-$U$ limit
Thus the factor $e^{\dag }$ in the numerator of (\ref{zz}) is canceled and we derive in the large-$U$ limit for the projected  operator of hole creation
\begin{equation}
\tilde{c}^{\dagger}_{i\sigma }={\sqrt{2}}\sum_{\sigma ^{\prime }}\hat{p}%
_{i\sigma\sigma ^{\prime }}f_{i\sigma }=\frac{1}{\sqrt{2}}\sum_{
^{\prime }\sigma}f_{i\sigma ^{\prime }}^{{}}[\delta _{\sigma \sigma ^{\prime
}}p_{i0}+(\mathbf{p}_{i}\mbox{\boldmath$\sigma $}_{\sigma ^{\prime }\sigma
})]  \label{eq:I.88}
\end{equation}%
or
\begin{eqnarray}
\tilde{c}^{\dagger}_{i\uparrow } &=&\frac{1}{\sqrt{2}}f_{i\uparrow
}(p_{i0}+p_{iz})+f_{i\downarrow }p_{i}^{+} \nonumber \\
\tilde{c}^{\dagger}_{i\downarrow } &=&\frac{1}{\sqrt{2}}f_{i\downarrow
}(p_{i0}-p_{iz})+f_{i\uparrow }p_{i}^{-}. \label{pr1}
\end{eqnarray}%
In particular, this representation satisfies exactly commutation relations for Hubbard's operators. However, the  multiplication rule, which is crucial for calculations, 
\begin{equation}
X_{i}(0,- )=X_{i}(0,+)X_{i}(+,- )
 \label{eq:6.21}
\end{equation}%
is satisfied only approximately ($\frac{1}{\sqrt{2}}(p_{i0}+p_{iz}) \simeq 1, X_{i}(+,-)\simeq p_{i}^{-}$).

In derivation of (\ref{pr1}), which was first performed in Ref. \cite{Irk},  spin-wave correction in matrices $L$ and $R$ were neglected to replace the operator $S^z$ by 1/2.
We can make the next step by noting that according to (\ref{s1}), (\ref{s2}), (\ref{con})
\begin{eqnarray}
\sqrt{1/2\pm S^z}\simeq \sqrt{1/2\pm 2p_{0}p_{z}/2}  \simeq (p_{0} \pm p_{z})/\sqrt{2}\end{eqnarray}
(a more strict derivation may be performed within path integral approach).
Then we can write the representation in terms of {\textit spin} operators,
\begin{equation}
\tilde{c}_{i \sigma }=\sum_{\sigma ^{\prime
	}}f^{\dagger}_{i\sigma ^{\prime }} [\frac{1}{{2}} \delta
	_{\sigma \sigma ^{\prime }}+(\mathbf{S}_i\mbox{\boldmath$\sigma $}_{\sigma
		^{\prime} \sigma} )] 
\label{eq:I.888}
\end{equation}%
or
\begin{eqnarray}
\tilde{c}_{i\uparrow } &=&f^{\dagger}_{i\uparrow
}(\frac{1}{{2}}+S_i^z)+f^{\dagger}_{i\downarrow }S_{i}^{+} \nonumber \\
\tilde{c}_{i\downarrow} &=&f^{\dagger}_{i\downarrow} (\frac{1}{{2}}-S_i^z)+f^{\dagger}_{i\uparrow }S_{i}^{-}. \label{pr11}
\end{eqnarray}
Although it  was justified above for small doping and
magnetic states only, it seems to be reasonable in a more general situation, as will be discussed below.

\subsection{Electron states and spin waves in the strongly correlated Hubbard model}

Now we can consider the electron and spin Green's functions for a saturated ferromagnet with the use of the Hamiltonian (\ref{H2}). In such a situation of small hole doping, the spin-up spinon states propagate freely and their band is almost half-filled, so that 
$$\tilde{n}_{\mathbf{k}}=\langle f_{\mathbf{k}\uparrow} f^\dag_{\mathbf{k}\uparrow} \rangle = n(t_{\mathbf{k}})$$
with $n(E)$ being the Fermi function.

The representation (\ref{pr11}) retains commutation relations for Hubbard's X-operators and even the multiplication rule (\ref{eq:6.21}), so that the calculations of electron and spin-wave spectra  can be performed step-by-step in analogy with \cite{FTT,FTT1} 
with using expansion in occupation numbers of holes and magnons.

Although the correlated electrons (holes), described by the operators (\ref{pr1}), (\ref{pr11}), are  composite particles,
the spin-up states propagate freely on the background of the ferromagnetic ordering, the temperature correction being proportional to $T^{5/2}$ owing to rotational invariance (however, the residue of the electron  Green's function has a more strong $T^{3/2}$ dependence). Physically, this free motion is due to condensation of  $p_{z}$-bosons.

On the other hand, the situation is quite non-trivial for spin-down states. Such a state is a complex of spinon $f_{\uparrow }^{\dag }$ and boson $%
(p^{-})^{\dag }$ 
so that in the simplest approximation we can write down the convolution of the spinon and magnon Green's functions to obtain

\begin{equation}
G_{\mathbf{k}\downarrow }^{0}(E)=\sum_{\mathbf{q}}\frac{\tilde{n}_{\mathbf{k}+\mathbf{q}}+N(\omega _{\mathbf{q}})}{E-t_{\mathbf{k}+\mathbf{q}}+\omega _{%
\mathbf{q}}}  \label{eq:I.779}
\end{equation}%
with $N(\omega )$ the Bose function. 
It should be noted that this result can be also reproduced starting from the boson representation (\ref{pr1}), Ref. \cite{Irk} and even in the more simple Schwinger boson representation (\ref{eq:6.31}), see Ref.\cite{Scr3}.
The instability of the saturated (half-metallic) state is described as condensation of these bosons.







To improve the approximation and describe the instability we  write down the equation of motion for  the Green's function
\begin{equation}
G_{\mathbf{k}\downarrow }(E)=
\langle\langle\tilde{c}_{{\bf k}\downarrow} 
|\tilde{c}^{\dagger}_{{\bf k}\downarrow} \rangle \rangle _{E} =
\sum_{\mathbf{q}}\Gamma _{\mathbf{kq}}(E),
\end{equation}
\begin{equation}
\Gamma _{\mathbf{kq}}(E)=\sum_{\mathbf{q^{\prime }}}\langle
\langle S_{\mathbf{q}}^{-}f_{\mathbf{\mathbf{q-k}\uparrow }}^{\dagger }|f_{\mathbf{q^{\prime }-k}\uparrow }S_{\mathbf{-q^{\prime }}}^{+}\rangle \rangle _{E}  \label{eq:I.778}
\end{equation}%
(note that the terms with $f_{\downarrow }$ do not work in low orders).
Commuting the operator $S_{\mathbf{q}}^{-}$ with the Hamiltonian (\ref{H2}) and performing decoupling we obtain for $T=0$ the equation for the Green's function in the
right-hand side of (\ref{eq:I.778})
\begin{eqnarray}
(E&-&t_{\mathbf{k}-\mathbf{q}}+\omega _{\mathbf{q}})\Gamma _{\mathbf{kq}%
}(E)\nonumber\\&=&\tilde{n}_{\mathbf{k} 
-\mathbf{q}}
[1-(t_{\mathbf{k}-\mathbf{p}}-t_{\mathbf{k}}) \sum_{\mathbf{p}}\Gamma _{\mathbf{k-p}}(E)].
\end{eqnarray}
The solution of this integral equation yields the result
\begin{equation}
G_{\mathbf{k}\downarrow }(E)=\left\{ E-t_{\mathbf{k}}+\left[ G_{\mathbf{k}%
\downarrow }^{0}(E)\right] ^{-1}\right\} ^{-1}  \label{eq:I.776}
\end{equation}%

The expressions (\ref{eq:I.779}) and (\ref{eq:I.776}) were previously derived using the many-electron approach of Hubbard's operators, as described in references \cite{FTT,FTT1}.
It was noted that these results bear resemblance to Anderson's spinons, which also exhibit zero residue in their Green's function. The Green's function (\ref{eq:I.779}) represents a purely non-quasiparticle state, indicating its unconventional nature. Due to the limited dependence on momentum ($\mathbf{k}$), the corresponding distribution function of these non-quasiparticle (incoherent) states exhibits low mobility and cannot provide electrical current.

Regarding the Green's function (\ref{eq:I.776}), when the doping level $1-n$ is small, it does not exhibit any poles below the Fermi level (for holes), confirming the previous conclusions. However, as the doping  increases, a spin-polaron pole $E_{F}$ emerges, resulting in the destruction of half-metallic ferromagnetism.

The description of the transition to the saturated state, where the spin-down quasiparticle residue diminishes, resembles that of the Mott transition in the paramagnetic Hubbard model \cite{Senthil2}. In this sense, the situation is somewhat comparable to a partial Mott transition occurring in the spin-down subband. For a more detailed discussion on this matter,  cf. the review \cite{Vojta} where the orbital-selective Mott transition is explored.

Now we calculate the correction to the magnon frequency. The equation of motion for the spin Green's function reads

\begin{equation}
(\omega - \omega_{{\bf q}} )
\langle \langle S^+_{{\bf q}} 
 |S^-_{{\bf -q}} \rangle \rangle _{\omega} =2\langle S^z\rangle +
\sum_{\mathbf{kp}}(t_{\mathbf{k}-\mathbf{q}}-t_{\mathbf{k}}) \Lambda _{\mathbf{kqp}%
}(\omega),
\end{equation}
\begin{equation}
\Lambda _{\mathbf{kqp}}(E)=\langle
\langle S_{\mathbf{k+q-p}}^{+}f_{\mathbf{p}\uparrow }f_{\mathbf{k}\uparrow }^{\dagger }|S^-_{\mathbf{-q}} \rangle \rangle _{\omega} \label{eq:I.77811}
\end{equation}%
In the same manner, we derive the integral equation
\begin{eqnarray}
(\omega-t_{\mathbf{k}}+t_{\mathbf{p}}-\omega _{\mathbf{k+q-p}})\Lambda _{\mathbf{kqp}
}(\omega)\nonumber \\
=\delta_{\mathbf{kp}}\tilde{n}_{\mathbf{k}}
+\sum_{\mathbf{r}}(t_{\mathbf{k+r+q-p}}-t_{\mathbf{r}}) \Lambda _{\mathbf{rqp}}(\omega). \label{Nag}
\end{eqnarray}
Neglecting the integral term (which is possible to leading order in the inverse nearest-neighbor number) we obtain  from the expansion of the Dyson equation the renormalized magnon frequency 
\begin{equation}
\Omega_{{\bf q}} =\omega_{{\bf q}} 
+\sum_{\mathbf{k}} (t_{\mathbf{k}-\mathbf{q}}-t_{\mathbf{k}})\tilde{n}_{\mathbf{k}}.
\end{equation}%
Th exact solution of Eq.(\ref{Nag}) provides accurate results to leading order in doping, in agreement with the consideration by Nagaoka \cite{Nagaoka} (see also \cite{FTT}).

\subsection{Antiferromagnetic case: small and large Fermi surfaces}

With increase of the doping, the Nagaoka ferromagnetic state becomes unstable. The instabilities can be also treated within the  Kotliar-Ruckenstein representation, as was performed numerically in   Refs.\cite{Fr,Ig}.
This representation, adopted above for the ferromagnetic phase (\ref{pr11}), is expected to hold also in the antiferromagnetic state when being written down in the local (rotating) coordinate system. 
Moreover, it will work also in the systems with strong spin fluctuations and short-range order 
(e.g., in the singlet RVB state), but not in the usual
structureless paramagnetic state.%




The representation (\ref{pr11}) is formally  very similar to the representation of the Fermi dopons $d^\dag_{i\sigma}$ \cite{Ribeiro,Punk} introduced to describe formation of small and large Fermi surfaces in doped two-dimensional cuprates. This has the form
\begin{equation}
	\tilde{c}_{i -\sigma }^{\dagger}=-\frac{\sigma}{\sqrt{2}}\sum_{\sigma ^{\prime
	}}d^{\dagger}_{i\sigma ^{\prime }}(1-n_{i-\sigma ^{\prime }}) [S\delta
	_{\sigma \sigma ^{\prime }}-(\mathbf{S}_i\mbox{\boldmath$\sigma $}_{\sigma
		^{\prime} \sigma} )].  \label{eq:I.78}
\end{equation}
where $\sigma=\pm$, $n_{i\sigma}=d^{\dagger}_{i\sigma}d_{i\sigma}$, and both Fermi spinon (Abrikosov) and Schwinger boson representations can be used for localized $S=1/2$ spins.
The latter representation has the advantage that hybridization of spinons with dopons can describe formation of the large Fermi surface including the localized states.


On the other, the Bose version \cite{Punk} can successfully describe  the small Fermi surface.
The presence of strong  antiferromagnetic 
correlations 
leads to that hopping of dopons between nearest-neighbors is strongly suppressed  owing to a local antiferromagnetic order \cite{Ribeiro,Punk}.
 Then small hole pockets of the Fermi surface, characteristic for the cuprates, are formed,  which tend to the ($\pi/2,\pi/2$) point of the Brillouin zone with increasing the short-range order \cite{Punk}. 

Thus we can apply our representation (\ref{pr11}) to the same problem.
Note that the description in terms of bosons $p$ (representation (\ref{pr1})) turns out to be oversimplified and incomplete, 
unlike the  approach bases on (\ref{pr11}), which provides description in terms of true spin degrees of freedom.


At first sight, the dopon representation can seem to be quite different from standard slave-boson representations.
However, the connection can be established by using the constraint $\sum_\sigma f_{i\sigma}^\dag f_{i\sigma}\simeq 1$ (which holds at small doping) and the Abrikosov representation for spin operators 
\begin{equation}
{S}_i^z = \frac{1}{2} (
f_{i\uparrow}^\dag f_{i\uparrow} - f_{i\downarrow}^\dag f_{i\downarrow} ), \,  {S}_i^\sigma = f_{i\sigma}^\dag f_{i-\sigma}.
\label{eq:I.780a}
\end{equation}
We rewrite (\ref{eq:I.78}) as
\begin{equation}
	\tilde{c}_{i\sigma}=\frac{1}{2}(d_{i\downarrow}^\dag f_{i\uparrow}^\dag
	-d_{i\uparrow}^\dag f_{i\downarrow}^\dag )f_{i\sigma}. \label{lz1}
\end{equation}
Then,  we can introduce Anderson's Bose holon operator as a singlet combination of Fermi spinon and new dopon  operators \cite{Ribeiro,Irk}, 
\begin{equation}
e_i=f_{i\uparrow}d_{i\downarrow}-f_{i	\downarrow}d_{i\uparrow}. 
\label{eq:I.78a}
\end{equation}
Thus, we return to Anderson's representation (\ref{eq:6.131}),
except for  the difference in the factor of $1/\sqrt{2}$.
The problem with this factor does not take place in our version of the Kotliar-Ruckenstein representation (\ref{pr11}) due to the factor of $M$ in (\ref{1z}).
Note that the dopon representation can be also derived in the many-electron approach of Hubbard's operators using the analogy with the equivalent narrow-band $s-d$ exchange model \cite{II,Scr}.

\section{Discussion}


We have demonstrated that the  Kotliar-Ruckenstein representation \cite{Kotliar86} provides unified description of paramagnetic and magnetic phases.
In the paramagnetic phase we present a new interpretation in terms of spin-charge separation and conservation of the Fermi surface in the insulator state.
We have also performed the derivation of the Hamiltonian in the magnetically ordered phase in the spin-wave region, which enables one to obtain an agreement with well-established results for ferromagnetic case.

The constructed approach is somewhat similar to the Holstein-Primakoff representation for Heisenberg systems.
The  Kotliar-Ruckenstein representation includes both Fermi and Bose (or spin) operators and has a rather complicated structure with radicals. Therefore, it in a sense solves the problem of describing transmutation statistics of auxiliary particles when passing from spin-liquid to magnetic phase, which was discussed in Sec. 2 and formulated earlier as an important issue (see, e.g., Ref.\cite{Punk1}).

Under deconfinement conditions, the characteristics of the energy spectrum are significantly affected by the presence of spinon excitations, and this should result in their  pronounced dependence on temperature on the scale of the Heisenberg interaction, which can be small in comparison with bare electron energies. The corresponding expressions for the Green's functions can be applied to write down the optical conductivity and describe the optical transitions between the Hubbard's subbands, as demonstrated in  Ref.\cite{raimondi}.

Anderson \cite{Anderson} applied the concept of spinons to explain the linear specific heat in copper-oxide systems by the contribution  of gapless spinons forming the Fermi surface  in the spin-liquid-like uniform resonating valence bonds (RVB) state. Although for the cuprates this point remains highly debatable,  there exist experimental evidences for contributions of spinons (gapless magnetic excitations)  to specific heat and thermal conductivity, etc.,  in some compounds with frustrated lattices (see, e.g., Refs. \cite{Lee,Shen,npj}),

At the same time, in magnetically ordered phase we have usual spin-wave excitations. These phases are also successfully described by the  Kotliar-Ruckenstein	 representation with account of incoherent states.
Exotic phases including both antiferromagnetic order and fractionalized  excitations (so-called AFM$^*$ or SDW$^*$ phase \cite{Vojta,Sachdev}) can be considered  too.
In systems with magnetic  or superconducting ground state, there is still a possibility for a spin-liquid-like state to emerge at intermediate temperatures, particularly in systems with frustration \cite{Sachdev}.

As we have demonstrated, topological transitions of a different nature with a reconstruction of the Fermi surface occur in antiferromagnetic and ferromagnetic \cite{Scr3} phases. 
It is evident now that the Mott transition leading to a non-magnetic ground state is closely linked to topological characteristics. This transition involves a deconfined spin-liquid state that exhibits fractionalization and extensive quantum entanglement \cite{Scr2}. Understanding the exotic correlated paramagnetic phase, which can possess intricate  structures, is a significant challenge in this context.

The author is grateful to Yu. N. Skryabin and M. I. Katsnelson for  useful discussions.
The research funding from the Ministry of Science and Higher Education of the Russian Federation (the state assignment, theme ``Quantum'' No. 122021000038-7)
is  acknowledged. The treatment of half-metallic ferromagnetic state is supported by the  grant of the Russian Science Foundation 23-42-00069.

\end{document}